\documentclass[a4paper,twoside]{article}
\newcommand{\affil}[1]{$^{\rm #1}$}
\usepackage[authoryear]{natbib}
\bibpunct{(}{)}{;}{a}{}{,}
\usepackage{graphicx}
\date{} 
\newcommand{\kms}{\mbox{km\,s$^{-1}$}}
\newcommand{\aap}{A\&A}

\newcommand{\apj}{ApJ}
\newcommand{\apjl}{ApJ}
\newcommand{\mnras}{MNRAS}

\newcommand{\aj}{AJ}

\def\kms{$\mbox{km s}^{-1}$}

\def\deg{^\circ}

\def\farcs{\hbox{$.\!\!^{\prime\prime}$}}
\def\arcs{^{\prime\prime}}
\def\mbh{$\mathrm{M_{BH}\ }$}
\def\mbulge{$\mathrm{M_{Bulge}\ }$}
\def\Msun{$\mathrm{M_{\odot}}$}

\title{\large\bf\flushleft The Supermassive Black Hole at the Heart of Centaurus~A: \\ Revealed by Gas- and Stellar Kinematics}
\author{\parbox{\textwidth}{\flushleft
\vspace{-0.5cm}
{\it Nadine Neumayer\affil{A,B}}\\
\vspace{0.4cm}
{\small \affil{A}\,European Southern Observatory, Karl-Schwarzschild-Str 2, 85748 Garching bei M\"unchen, Germany}\\
{\small \affil{B}\,Email: nneumaye@eso.org}}}
\begin{document}
\twocolumn[
\begin{changemargin}{.8cm}{.5cm}
\begin{minipage}{.9\textwidth}
\vspace{-1cm}
\maketitle
%
%
\small{\bf Abstract:}
At less than 4~Mpc distance the radio galaxy NGC~5128 (Centaurus~A) is the prime example to study the supermassive black hole and its influence on the environment in great detail. 
To model and understand the feeding and feedback mechanisms one needs an accurate determination of the mass of the supermassive black hole. 
The aim of this review is to give an overview of the recent studies that have been dedicated to measure the black hole mass in Centaurus~A from both gas and stellar kinematics. It shows how the advancement in observing techniques and instrumentation drive the field of black hole mass measurements and concludes that adaptive optics assisted integral field spectroscopy is the key to identify the effects of the AGN on the surrounding ionised gas. Using data from SINFONI at the ESO Very Large Telescope, the best-fit black hole mass is \mbh$=4.5^{+1.7}_{-1.0} \times 10^7$\Msun (from H$_2$ kinematics) and  \mbh$= (5.5\pm 3.0)\times 10^7$\Msun (from stellar kinematics; both with 3$\sigma$ errors). This is one of the cleanest gas vs star comparison of a \mbh determination, and brings Centaurus~A into agreement with the \mbh-$\sigma$ relation.


\medskip
\medskip
\end{minipage}
\end{changemargin}
]
\small

\section{Introduction}
Centaurus~A (hereafter Cen~A) is the posterchild of an active galaxy. Also known as NGC~5128 it is the closest elliptical galaxy, the closest recent merger, and hosts the closest active galactic nucleus (AGN). The distance to Cen~A has long been under debate and seems to converge to the value 3.8~Mpc \citep{rejkuba04,karachentsev07,harris09}.

Cen~A is one of the most studied galaxies in the sky and its proximity makes it an ideal candidate to test AGN models, the connection of merging and star formation, and the influence of the jet.
Its proximity also makes it one of the prime targets to reliably measure the mass of the supermassive black hole, suspected to hide behind the obscuring dust lane. This prominent dust lane has hindered the study of CenA's black hole with optical HST spectroscopy as was common practice for other galaxies \citep[e.g. M87,][]{macchetto97}, and requires to move to the near-infrared wavelength for a detailed study of its properties.\\
Interestingly, the first measurements of Cen~A's black hole mass \citep{marconi01,silge05} brought it almost a factor of ten above the so called M-$\sigma$ relation \citep{ferrarese00,gebhardt00}, and made Cen~A one of the largest outlier to this relation. It was not clear, whether this offset to the M-$\sigma$ relation is intrinsic to Cen~A, maybe due to the presence of an AGN or its merger history, or whether it can be accounted for by the fact that seeing limited studies do not fully resolve the sphere of influence of the black hole, which would be at $\sim 0\farcs4$, taking the predicted mass from the M-$\sigma$ relation \citep{ferrarese00,gebhardt00,tremaine02} for granted.\\

There are several methods to detect black holes in galaxy centres and to measure their masses. For nearby galaxies the most direct methods take gas and stars as kinematics traces of the nuclear region. Dynamical models are then constructed to explain the kinematics.  These models rely on a couple of assumptions (circular gravitational motion of the gas, superposition of orbits for the stars). To test the reliability of both approaches it is important to compare their model results.
In general, the gas kinematical method has the advantage of using a relatively simple modelling approach and dealing with relatively short exposure times. However, the gas kinematics might be influenced by non-gravitational motions that bias or even falsify the method. Moreover, not every galaxy nucleus has detectable gas emission lines. The stellar dynamical approach has the clear advantage that stars are always present in galaxy nuclei and that their motion is always gravitational. However, the derivation of stellar kinematics requires long exposure times and gets very complicated in the close vicinity of an AGN, as the stellar absorption lines are diluted by the AGN continuum radiation. In addition, stellar dynamical models are very complex leading to potential indeterminacy \citep[e.g.][]{cretton04,valluri04}.
Centaurus~A is a prime example for testing both modelling approaches on our doorstep. It allows both gas and stellar kinematical studies at a spatial resolution of $\sim 2$pc  (corresponding to $0\farcs1$) using either 
the HST or ground based adaptive optics instrumentation.\\

This paper reviews the status of the measurement of the black hole mass in Centaurus~A and shows how the advance in instrumentation is driving the field of black hole mass modelling.  Section \ref{studies} presents the different studies dedicated to the black hole mass measurement in Cen~A, while Section \ref{lessons} states what has been learned from these studies. Finally, Section \ref{conc} summarises the observational facts for Cen~A's black hole. All studies on Cen~A's black hole mass assumed a distance to Cen~A of D=3.5~Mpc. The derived black hole mass directly relates to the adopted distance (\mbh$\propto$ D)  and so all results need to be increased by 8.5\% given the distance measurement of 3.8~Mpc recommended by \cite{harris09}.

\section{Dedicated Studies\label{studies}}

This section briefly lists and describes the studies that were performed to get the most accurate black hole mass determination to date of the supermassive black hole in Centaurus~A. Each of these papers sets the state-of-the-art observations and modelling techniques available at the time the study was pursued. Gas and stellar kinematical studies are divided into two groups for comparative purposes. For details concerning the individual studies the reader is referred to the original papers.

\subsection{Gas kinematical studies}

\subsubsection*{Marconi et al. 2001\\ Peering through the dust}
An extensive imaging study with WFPC2 and NICMOS on board the {\it Hubble Space Telescope (HST)} (PI: E. Schreier) has revealed the nucleus of Cen~A in near-IR (K and H) and, for the first time, in the optical (I and V) at unprecedented spatial resolution \citep{marconi00}. \cite{schreier98} detected an elongated structure in Pa$\alpha$ along the jet position angle and interpreted it as a thin gas disc.  \cite{marconi01} used the near-infrared spectrograph and imager ISAAC \citep{moorwood99} on the ESO Very Large Telescope to retrieve spectra in J, H, and K$_{\mathrm s}$ Band under excellent seeing conditions of 0\farcs5 - 0\farcs6. They used the HST narrow band images as a reference to position the slit along three position angles and derived the kinematics for both ionised and molecular gas from their emission lines (Pa$\beta$, [FeII], Br$\gamma$, and H$_2$). The ionised gas kinematics are modelled by a thin disc model in the gravitational potential of a point mass, representing the black hole. Free parameters in the model are the inclination angle of the modelled disc, the black hole mass and the systemic velocity of the system. The inclination angle and the central mass are strongly coupled, since the amplitude of the rotation curve is proportional to $\sqrt{\mathrm{M_{BH}}} \times \sin(i)$. The ISAAC data do reject disc inclinations $>60\deg$, but do not constrain the disc inclination further. This gets reflected in the large uncertainties of the black home mass determination. The best fit value is \mbh$= 2^{+3.0}_{-1.4} \times 10^8$\Msun.

\subsubsection*{H\"aring-Neumayer et al. 2006\\ Adaptive Optics works!}
\begin{figure}[h]
\begin{center}
\includegraphics[scale=0.4, angle=0]{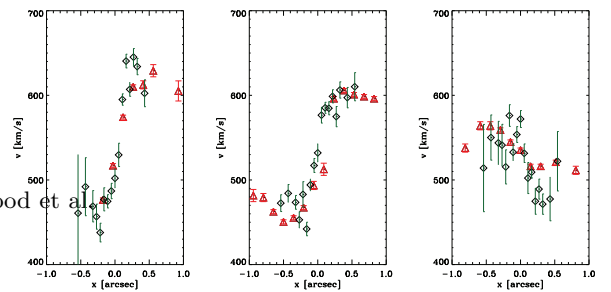}
\caption{Velocity curves extracted from the [FeII] emission line at $\lambda1.644 \,  \mu$m from NACO (diamonds) and ISAAC data (triangles), along three slit positions centred on the nucleus of Cen~A (P.A.= $-44.5\deg$, $32.5\deg$, and $82.5\deg$ from left to right). The NACO data have almost factor of five better resolution than the ISAAC data and are overall consistent.}\label{compare_isaac_naco}
\end{center}
\end{figure}

In the year 2004 the resolving power of ground based instrumentation has taken a big leap ahead when the first adaptive optics systems went online. Using NAOS-CONICA (NACO) \citep{rousset03,lenzen98} at the ESO VLT, \citet{hn06} (hereafter HN+06)
followed the approach of \cite{marconi01} and used long-slit spectra at the same position angles to probe the kinematics of the ionised gas. Due to the correction of the atmospheric distortion, the spatial resolution of the spectra increased by almost a factor of five compared to the study of \cite{marconi01} and provided images and spectra at (or almost at) the diffraction limit of the VLT ($0\farcs06$).  The data points are shown in Figure~\ref{compare_isaac_naco} along with the [FeII] kinematics derived by \cite{marconi01} from ISAAC seeing limited data. HN+06 note that the [FeII] velocity dispersion is very high along all slit positions and cannot be fully accounted for by unresolved rotation. They have thus included a pressure term in their gas kinematical model. Moreover, they take into account the underlying potential of the stars (that was not accounted for by \citet{marconi01}) and model the [FeII] kinematics in the joint potential of the stars and an additional point mass, assumed to be the black hole. Their data along four long-slit position angles does also not constrain the inclination angle of the modelled gas disc well. Their best fitting model gives $i=45\deg$ and \mbh$= 6.1^{+0.6}_{-0.8} \times 10^7$\Msun. The decrease in black hole mass compared to \cite{marconi01} is due to (i) accounting for the stellar potential, (ii) the higher resolution of the data, and (iii) the higher inclination angle of the gas disc.

\subsubsection*{Marconi et al. 2006\\ Pushing STIS to the reddest end}
\begin{figure}[h]
\begin{center}
\includegraphics[scale=0.9, angle=0]{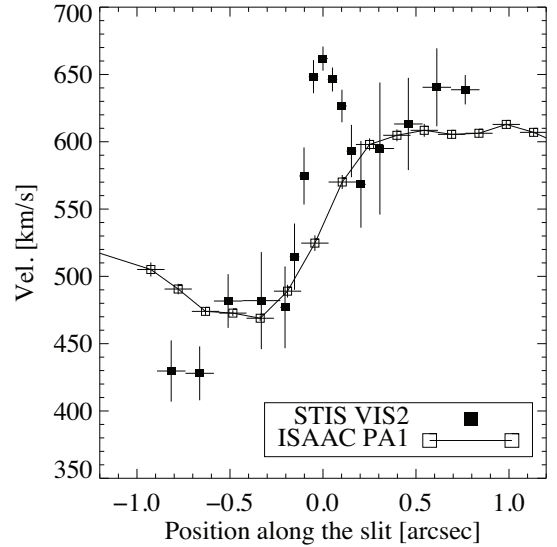}
\caption{Comparison of average velocities from STIS derived from [SIII] emission and
ISAAC derived from [FeII] and Pa$\alpha$, showing the increase in spatial resolution of almost a factor of five. This is a reproduction of Fig. 8 from \cite{marconi06}}\label{compare_isaac_stis}
\end{center}
\end{figure}

Knowing exactly where to look, \cite{marconi06} have pointed the HST spectrograph STIS on the dust enshrouded nucleus, that is not visible in the optical wavelengths. Pushing STIS to the reddest limit, they used the [SIII] $\lambda9533 \, {\mathrm \AA}$ line to study the
kinematics of the ionized gas in the nuclear region with a $0\farcs1$ spatial resolution. 
The STIS data were analysed in conjunction with the ground-based near-infrared VLT ISAAC spectra used by \cite{marconi01}, and have a gain in spatial resolution of almost a factor of five (see Figure~\ref{compare_isaac_stis}). The gas kinematical analysis provides a mass of \mbh\, $ = (1.1\pm 0.1) \times 10^8$\Msun\, for an assumed disc 
inclination of $ i=25\deg$ and \mbh$=(6.5 \pm 0.7) \times 10^7 $\Msun\, for $i=35\deg$, the largest inclination value allowed by their data.
In this study, \cite{marconi06} did extensive tests on the influence of the choice of the parametrisation of the surface brightness profile of the assumed gas disc, that is used to luminosity-weight the modelled emission lines and to resemble the observations. They found that the associated systematic errors are no larger than 0.08 in log\mbh comparable
with statistical errors and indicating that the method is robust. However, the intrinsic surface brightness distribution has a large
impact on the value of the gas velocity dispersion. They conclude that a mismatch between the observed and modelled velocity dispersion is not
necessarily an indication of non-circular motions or kinematically hot gas, but is as easily due to an inaccurate computation
arising from too course a model grid, or the adoption of an intrinsic brightness distribution which is too smooth.
Therefore, they did not include pressure support in their model. The stellar potential is accounted for by the model.
The spatial resolution of the data is $\sim 0\farcs1$ the same as in HN+06 and both studies use ionised gas as the kinematic tracer ([SIII] and [FeII], respectively). The largest discrepancy between the two studies is the modelled disc inclination angle. \cite{marconi06} find that the inclination angle is $i\le35\deg$, while HN+06 find their best model at $i=45\deg$ and exclude smaller inclination angles on the basis of the jet inclination angle (assuming a orthogonal jet-disc geometry and a jet inclination angle of $\sim 50\deg$ \citep{tingay98,hardcastle03}). The inclination angle remains the strongest source of uncertainty in the model. The results of \cite{marconi06} and HN+06 are in very good agreement, and also agree with \cite{marconi01} inside the error-bars. 

\subsubsection*{Krajnovi\'c, Sharp \& Thatte 2007\\ Integral Field Spectroscopy to the rescue!}
All previous approaches were restricted to a limited number of slit position angles (3 for \cite{marconi01}, 4 for HN+06, and 2 for \cite{marconi06}) and did not allow to trace the kinematics of the gas outside these stripes. This changed when \cite{krajnovic07} analysed the data from the Cambridge Infrared Panoramic
Survey Spectrograph \citep[CIRPASS,][]{parry04}, an integral field spectrograph mounted on the Gemini South telescope. They detect two dimensional distributions of the following emission lines: [PII], [FeII], and Pa$\beta$, and extracted spatially resolved 2D kinematics from Pa$\beta$ and [FeII]. All emission-line regions are part of the same kinematic structure which shows a twist in the zero-velocity curve beyond $\sim 1\arcs$ (for both Pa$\beta$ and [Fe II]). The kinematics of the two emission lines are similar, but the Pa$\beta$ velocity
gradient is steeper in the centre while the velocity dispersion is low everywhere. The velocity
dispersion of the [FeII] emission is relatively high featuring a plateau, approximately oriented
in the same way as the central part of the warped disc. \cite{krajnovic07} use the 2D kinematic information to test
the hypothesis that the ionized gas is distributed in a circularly rotating disc. Assuming simple
disc geometry they estimate the mass of the central black hole using Pa$\beta$ kinematics, which
is consistent with being distributed in a circularly rotating disc. Their best fit model gives \mbh $=8.25^{+2.25}_{-4.25} \times 10^7$\Msun, for PA $=-3\deg$ and $i=25\deg$, and a stellar mass-to-light ratio of 0.0 $\mathrm{M/L_{\odot}}$, indicating that the Pa$\beta$ gas disc is fully under the gravitational influence of the black hole.
To test their model against previous gas dynamical models, \cite{krajnovic07} constructed models with different geometrical orientations taken from previous studies. They find that these give similar results, and are often statistically indistinguishable. However, the mass predicted by the M-$\sigma$ relation \citep{ferrarese00,gebhardt00} is excluded by all their models.

\subsubsection*{Neumayer et al. 2007\\ Integral Field Spectroscopy combined with Adaptive Optics}
\begin{figure}[h]
\begin{center}
\includegraphics[scale=1., angle=0]{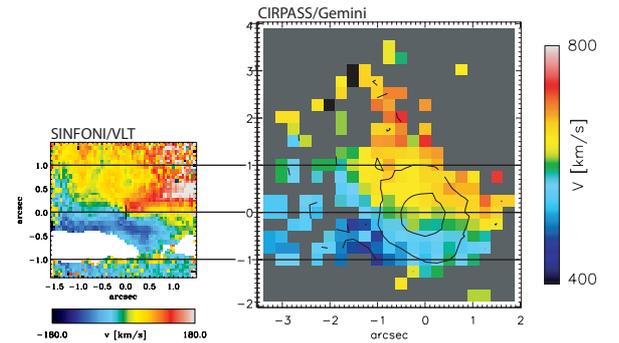}
\caption{The [FeII] velocity maps compared from SINFONI adaptive optics and CIRPASS seeing limited data. The Figures are taken from \cite{neumayer07} and \cite{krajnovic07} respectively. The overall kinematic structure of the two maps is in good agreement. However, the adaptive optics assisted SINFONI map shows detailed kinematic substructures and reveals the influence of the jet on the kinematics.}\label{sinfoni_cirpass}
\end{center}
\end{figure}

Another advance in observing techinque allowed a big jump ahead in the studies of black hole mass modelling. Combining the resolving capabilities of adaptive optics with the power of integral field spectroscopy, makes SINFONI on the VLT \citep{eisenhauer03,bonnet04} the perfect instrument for this case. SINFONI works in the near-infrared and allows to peer through the dust lane hiding Cen~A's nucleus. The adaptive optics assisted data presented by \cite{neumayer07} have a spatial resolution of $0\farcs12$ and cover the central $3\arcs \times 3\arcs$. The data cubes in H- and K-band exhibit a wealth of ionised gas emission lines: [FeII], [SiVI], HeI, Br$\gamma$, [CaVIII], and several transitions of H$_2$. In addition the data enable the extraction of stellar kinematics \citep[which will be presented in detail in Section \ref{cap} ;][]{cappellari09}. 
The high spatial resolution two-dimensional maps of the ionised gas show clearly two distinct kinematic features: rotation and transverse motion (see Figure~\ref{sinfoni_cirpass}). The transverse motion is oriented along the jet direction (PA=55$\deg$) and is strongest for the high-ionisation lines ([SiVI] and [CaVIII]).  However, also the tracers used in previous studies like [FeII] and Br$\gamma$ are influenced by this non-gravitational motion that has not been accounted for by the dynamical models. Long-slit spectroscopy was not able to spot this additional kinematic feature and therefore leads to an over-simplified picture, given the assumption that the gas moves in a circularly rotating disc, which underlies the gas dynamical models.\\
The picture is surprisingly different when looking at the kinematics of molecular hydrogen rather than the ionised gas. This kinematic tracer shows no kinematic component along the jet direction and seems to be dominated by gravitational motion only. Therefore, \cite{neumayer07} have based their dynamical model on these kinematics.
The two dimensional kinematic maps constrain the geometry of the disc much better than the few long-slit position angles have done before. The disc is not flat, but appears warped. The disc parameters, i.e. the inclination angle and the major axis position angle, were fitted using the kinemetry software by \cite{krajnovic06}. The fitted mean inclination angle of the H$_2$ gas disc is $45\deg\pm7\deg$ and the mean position angle of the major axis is 155$\deg$. The inclination angle is in very good agreement with the value derived by \cite{hardcastle03} from VLA data ($20\deg < i < 50\deg$), it is, however, somewhat smaller than the value from VLBI data derived by \cite{tingay98} ($50\deg < i < 80\deg$). This well constrained disc parameters enable a significant reduction of the uncertainty of the modelled black hole mass. The assumed mass-to-light ratio of the underlying stellar body, M/L$_K = (0.72 \pm 0.04) M_{\odot}/L_{\odot}$, is taken from the stellar kinematical model by \cite{silge05} and confirmed by \cite{cappellari09}. \cite{neumayer07} modelled the H$_2$ kinematics on tilted rings following the kinemetry model, keeping the relative inclination angle between the rings fixed but varying the overall inclination. The overall best-fit black hole mass is  \mbh $=4.5^{+1.7}_{-1.0} \times 10^7$\Msun and the best fit mean inclination angle is $34\deg\pm4\deg$ (3 $\sigma$ error bars).

\subsection{Stellar kinematical studies}

\subsubsection*{Silge et al. 2005\\ Near-Infrared long-slit spectra}
Using the Gemini Near-Infrared Spectrograph (GNIRS) on Gemini South, \cite{silge05} measured the stellar kinematics under very good natural seeing conditions of $0\farcs45$ and $0\farcs6$. They put the slit along two position  angles across the nucleus of Cen~A, centred on the AGN. The first observation was made with the slit oriented perpendicular to the inner dust ring (along the major axis at large radii), while the second observation was made with the slit oriented parallel to the dust lane but offset from the centre by $0\farcs85$.  
The stellar kinematics are extracted using the region around the CO band heads at $2.3\, \mu$m.
The central spatial bins of the first observations are dominated by emission from the AGN and the CO band heads are diluted by this emission. To recover the stellar signatures inside the central $0\farcs3$, \cite{silge05} have removed the AGN emission by measuring the AGN contribution by its dilution of the equivalent width of the $(2-1)^{12}$CO band head, assuming that the equivalent width is largely constant with radius in this galaxy for regions outside the AGN. 
After subtracting the AGN contribution, the stellar kinematics were extracted from the spectra by fitting a stellar template spectrum (convolved with a velocity profile) to the observations. The choice of template star is important in order to avoid mismatch \citep{silge03} and so a variety of template stellar spectra is simultaneously fitted for the velocity profile and the stellar template weights. As a result the fitting procedure provides the line-of-sight velocity distribution (in a non-parametric form) as well as the stellar population information.
The luminosity weighted stellar velocity dispersion along the slit parallel to the dust disc is $138\pm10$ \kms.\\
The black hole mass is determined using axisymmetric orbit-based models. The full line-of-sight velocity distribution is fitted to the data and the best fit parameters, stellar M/L and \mbh, are derived via a $\chi^2$ minimisation (the interested reader is referred to \cite{silge05} for details). From the long slit data along two slit positions alone it is not possible to identify the rotation axis and so the modelling procedure was performed twice: first matching the rotation axis to the dusk disc and second matching the rotation axis to the photometric major axis and non-rotating kinematics along the dust disc. It turns out that the two assumptions give very similar black hole masses but the second approach results in significantly lower $\chi^2$ values.\\
Although NGC~5128 appears very round, its actual shape is unknown and it could be intrinsically quite flattened. Therefore, \cite{silge05} repeated their models for three different intrinsic shapes (i.e intrinsic axis ratios) of Cen~A:
intrinsically round (1:1:1), at an inclination of $i=20\deg$ (1:1:0.5), and $i=45\deg$ (1:1:0.9). They find that this drastic change in flattening does not change the black hole mass much. The $\chi^2$ of the intrinsically round model is significantly less than either flattened model, and so this is adopted as the best-fit model with \mbh$=2.4^{+0.3}_{-0.2}\times 10^8$\Msun and stellar M/L$_{K}=0.68^{+0.01}_{-0.02}$ in solar units.

\subsubsection*{Cappellari et al. 2009\\ Integral field spectroscopy combined with adaptive optics\label{cap}}
Using the adaptive optics assisted integral field spectrograph SINFONI at the ESO Very Large Telescope, \cite{cappellari09} extracted the stellar kinematics from the nuclear region of NGC~5128 with a spatial resolution of $0\farcs17$. This means an increase in spatial resolution by a factor of three to four compared to the seeing limited data of \cite{silge05}. The high S/N data cover the central $8\arcs \times 8\arcs$ and allow the extraction of the stellar line-of-sight velocity distribution, which is given in parametric form of a Gauss-Hermite expansion.  At radii where the nucleus dominates, the stellar absorptions are diluted, resulting in a strong decrease in the observed line-strength $\gamma$. An inaccurate modelling of the non-thermal dilution can cause large errors in the measured velocity dispersion $\sigma$. Similarly, the rise in the non-thermal continuum could be incorrectly interpreted as a variation of the stellar population, requiring a change in the stellar template mix and also producing an error in $\sigma$. As there is no evidence for a sudden change in the population in the nucleus of Cen~A, the authors argue the safest choice is to assume that the stellar template is fixed and to model the non-thermal continuum via additive polynomials. This approach allows the $\sigma$ to be reliably extracted in the high-resolution observations down to $\ge 0\farcs2$.
For the first time, a low-level clear sense of stellar rotation is detected for Cen~A which is counter-rotating with respect to the gas. The global kinematical major axis is at PA$_{\mathrm kin} = 167\pm8\deg$ which is in between the dust disc and the galaxy's photometric major axis, and therefore between the two position angles that \cite{silge05} assumed for their kinematic models. The derived systemic velocity is v$_{\mathrm syst}= 531 \pm 5$ \kms.\\

\begin{figure}[h]
\begin{center}
\includegraphics[scale=0.8, angle=0]{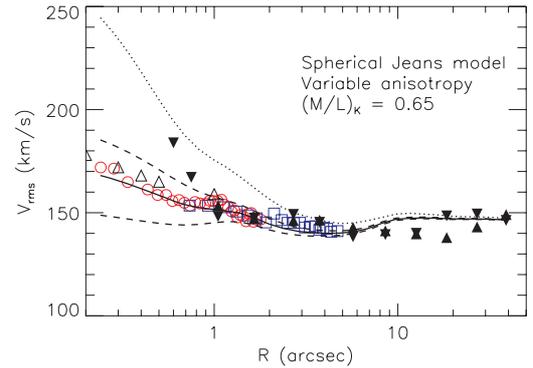}
\caption{Spherical anisotropic Jeans model to check the \mbh determinations. The root mean square stellar velocity, $\mathrm{v_{rms} = \sqrt{v^2+\sigma^2}}$, is plotted as a function of radius from the centre of Cen~A, derived from SINFONI observations (100~mas, red open circles; 250~mas scale, blue open squares) and from GNIRS observations of \cite{silge05} along the major axis (filled downward
triangle) and the minor axis (filled upward triangle). The discrepancy between the SINFONI and GNIRS data inside $1\arcs$ is due to a different subtraction of the AGN continuum. The solid
line shows the prediction of an anisotropic Jeans model, having the best-fitting \mbh and
M/L derived from the Schwarzschild model, the dashed lines represent the upper and lower 3$\sigma$ errors in the Schwarzschild model. The dotted lines has \mbh$=2\times 10^8$\Msun, corresponding to the best-fitting value of \cite{silge05}. This figure is reproduced from \cite{cappellari09}. }\label{silge_cappellari}
\end{center}
\end{figure}

Having in mind the complex dynamical and morphological structure that Cen~A possesses, 
\cite{cappellari09} constructed two types of models to test the
sensitivity of the \mbh estimate to the assumed geometry and dynamics; (i) an axisymmetric orbit-based model, to reproduce
in detail the kinematic observations and (ii) a simple anisotropic spherical Jeans model to qualitatively check the \mbh determination. This test gave very consistent results and is shown in Figure~\ref{silge_cappellari} along with a comparison to the data presented by \cite{silge05}. The best-fitting 
value for the black hole mass from the Schwarzschild model is \mbh$= (5.5\pm 3.0)\times 10^7$\Msun (3$\sigma$ errors) and the corresponding best fitting stellar M/L$_{K}$ is $(0.65\pm0.15) \mathrm{M/L_{\odot}}$.

\section{Lessons learned \label{lessons}}
The ever increasing accuracy of the black hole mass measurement in Centaurus~A over the past eight years reflects the advancement in observing techniques, especially in the near infrared. High spatial resolution observations are crucial to determine the influence of the central black hole on the gas and stellar kinematics. Even in an object that is as nearby as Cen~A, where the sphere of influence of the black hole is comfortably resolved ($\mathrm{R_{BH}}\sim 0\farcs7$, using the mass measurement of \citet{cappellari09}) using HST or adaptive optics observations.\\
Adaptive optics actually works! With the advent of near-infrared adaptive optics assisted spectrographs, such as NACO at the VLT, even dust enshrouded  galaxy nuclei became accessible at a spatial resolution of $0\farcs1$. This gets even more powerful when combined with integral-field spectroscopy. Mapping the gas and the stars in 3D allows to directly and simultaneously compare the morphology and kinematics of different gas species plus the stars. Having this powerful tool in hand, the influence of the inner jet on the kinematics of the ionised gas in Cen~A could be revealed, and moreover, molecular hydrogen could be identified as the ideal gas tracer for the central gravitational potential. The physical state of the gas is therefore very important when using it as a tracer for the dynamical models. The decrease of the best-fit value from \cite{marconi01} to \cite{marconi06} and HN+06 was due to the increase in spatial resolution plus taking into account the stellar contribution to the potential, while the decrease in \mbh from \cite{marconi06} and HN+06 to \cite{neumayer07} was due to the fact that the kinematic tracer changed from ionised gas to molecular hydrogen.\\
The presence of an AGN can definitely influence the kinematics of the gas, while the stellar kinematics should be unchanged by this. However, the extraction of the stellar kinematics from the spectral absorption features gets increasingly difficult in the close vicinity of the AGN, as the non-thermal continuum dilutes the stellar absorption lines and needs to be accounted for in the extraction process. This is the main difference in the analysis of \cite{silge05} and \cite{cappellari09}. While \cite{silge05} first subtract the AGN contribution and then fit the stellar line-of-sight-velocity distribution, \cite{cappellari09} include the fit of the AGN continuum in the the extraction of the stellar kinematics. To get reliable kinematics high signal-to-noise data are required. This is a very interesting lesson that we learned from Cen~A, and we should be cautious when extracting kinematics from other, more distant objects. Cen~A is indeed the closest AGN and at the same time it is very complex. Every leap in instrumentation development is likely to reveal more complex substructures. This warrants our continuous attention, in order to reveal intrinsic properties in the data and understand shortcomings in the models that aim to predict the observations.\\

\begin{figure*}[h]
\begin{center}
\includegraphics[scale=0.8, angle=0]{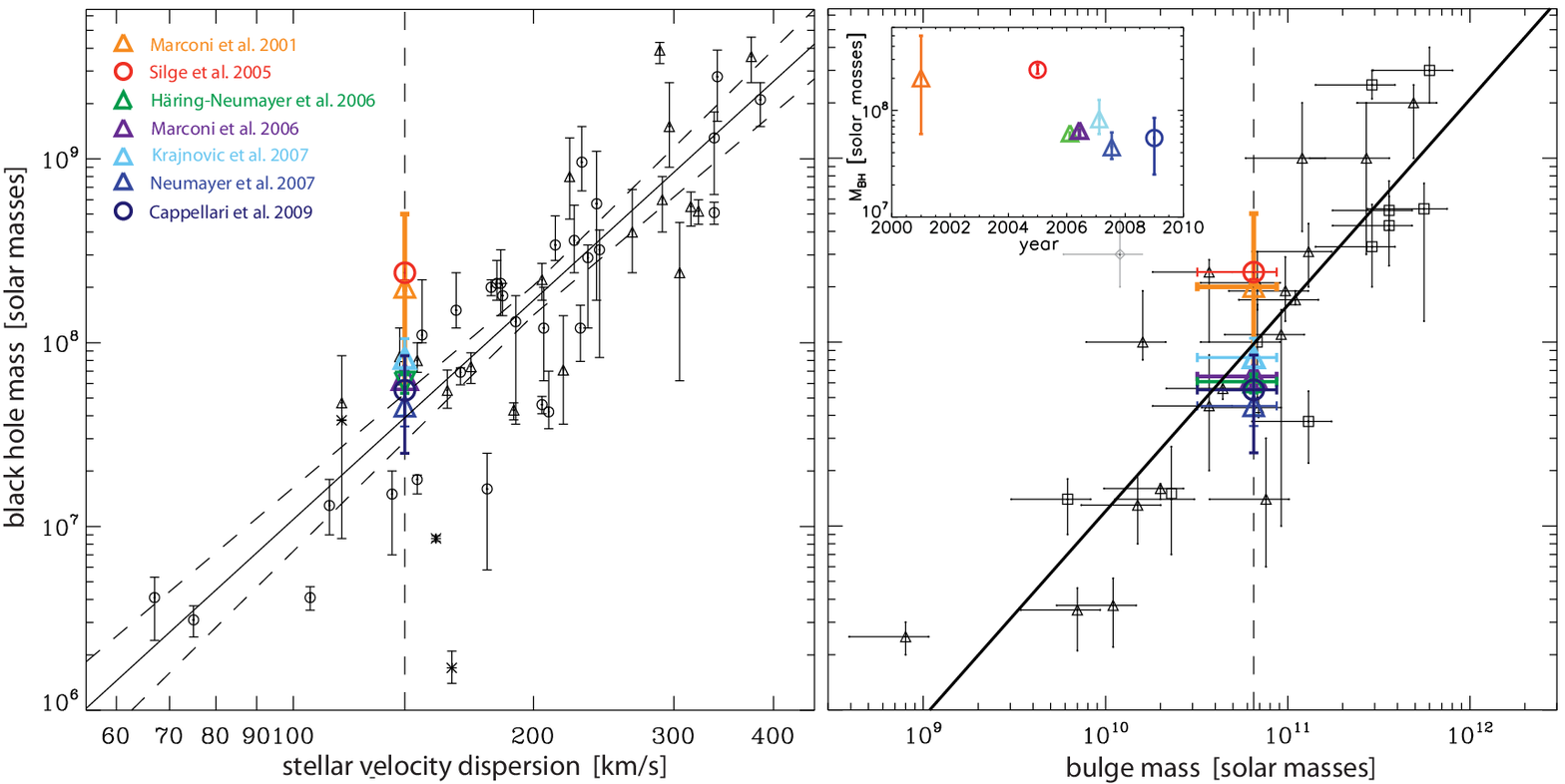}
\caption{Cen~A's black hole mass measurements are plotted on top of the black hole mass galaxy scaling relations. The left panel shows the \mbh-$\sigma$ relation reproduced after \cite{gultekin09}, with the overplotted line being the best fit relation to the elliptical galaxy subsample of all plotted galaxies. The right panel shows the \mbh-\mbulge relation as presented in \cite{haering04} with Cen~A values over-plotted. The Cen~A plot symbols as well as their time sequence are indicated in the upper left corner of the left and right panels, respectively. Values published in the same year are separated horizontally for clarity. In the left panel (and for Cen~A measurements) triangles refer to gas kinematical measurements, while circles refer to dynamical models using stellar kinematics.}\label{relations}
\end{center}
\end{figure*}

\section{Summary \label{conc}}
In principle all presented \mbh determinations for Cen~A using gas kinematical models agree within the error bars. The only significant disagreement in \mbh is with the previous stellar dynamical \mbh determination by \cite{silge05} and subsequent measurements \citep[HN+06, ][]{marconi06,krajnovic07,neumayer07,cappellari09}. The disagreement is likely caused by a difference in the data quality and in the treatment of the contribution from the central non-thermal continuum in the kinematic extraction \citep[compared to][]{cappellari09}. It is not due to differences in the details of the modelling methods. \\
The latest mass determination for the black hole at the nucleus of Centaurus~A, using high resolution integral field stellar kinematics, gives \mbh$=(5.5\pm3.0)\times 10^7$\Msun \citep{cappellari09} and agrees very well with the value \mbh$=4.5^{+1.7}_{-1.0} \times 10^7$\Msun determined using $H_2$ gas kinematics from the same SINFONI data \citep{neumayer07} . 
This gas versus stars \mbh com-parison constitutes one of the most robust and accurate ones, due to a very well resolved black hole sphere of influence ($R_{\mathrm BH}\sim 0\farcs7$ compared to a PSF FWHM$\sim 0\farcs12$) and thanks to the use of high-resolution integral-field data for both kinematical tracers. \\
Comparable and equally successful comparisons were done by \cite{shapiro06} and \cite{siopis09}, while a less good agreement was found in \cite{cappellari1459}, likely due to the disturbed gas kinematics.
There are certainly weaknesses in both the stellar and gas dynamical models, as e.g. the assumption of circular gravitational motions for gas kinematics and the complexity for the modelling of stellar kinematics. However, the agreement between the two modelling approaches for a growing number of cases strengthens the reliability of both methods.\\

 With the latest gas and stars \mbh determination Cen~A lies within the errors on the \mbh$-\sigma$ relation as given by either \cite{tremaine02}, \cite{ferrarese05}, or \cite{gultekin09}, and is in agreement with the relation of black hole mass to bulge mass as given by \cite{haering04} (see Figure~\ref{relations}). As the value of Cen~A's black hole mass has changed since 2001, so has the \mbh$-\sigma$ relation. The scatter in the relation seems to depend on the Hubble type of the chosen galaxies and is smallest for elliptical galaxies \citep{hu08,graham08,gultekin09}. The latest SINFONI \mbh measurement is in agreement with the relation for ellipticals (Figure~5).

\section*{Acknowledgments} 
I thank the DFG cluster of excellence 'Origin and Structure of the Universe' for financial support, and Davor Krajnovi\'c for a critical reading of the manuscript. Moreover, I would like to thank the science organizing committee for the invitation to the Many Faces of Centaurus~A Conference in Sydney as well as for financial support. Finally, I thank the anonymous referee, who helped to improve the manuscript.



\end{document}